\documentclass[twoside,letterpaper,12pt]{article}
\pdfoutput=1
\usepackage{authblk}

\usepackage[english]{babel}
\usepackage{color}
\usepackage{multirow}
\usepackage{multicol}
\usepackage{subfigure}
\usepackage{graphicx}
\usepackage{epsfig}
\usepackage{amsmath}
\usepackage{amssymb}
\usepackage{amsfonts}
\usepackage{wasysym}
\usepackage{slashed}
\usepackage{bbm}
\usepackage{ctable,longtable}
\usepackage{cancel}
\usepackage{txfonts}

\usepackage[colorlinks=true,urlcolor=black,linkcolor=black,citecolor=black,bookmarks=true]{hyperref}

\def\SU2U1{{\rm SU}(2)\times{\rm U}(1)}

\mathchardef\qsm=63
\mathchardef\pls=43
\mathchardef\mns=512
\mathchardef\plm=518
\mathchardef\eql=61
\mathchardef\smallleft=300
\mathchardef\smallright=301
\mathchardef\perslsh=47
\mathchardef\les=316
\mathchardef\gre=318
\mathchardef\leq=532
\mathchardef\grq=533
\chardef\usc=95
\chardef\til=126

\def\sqr#1#2#3{{\vcenter{\hrule height.#3ex\hbox{\vrule width.#2ex height#1ex
    \kern#1ex\vrule width.#3ex}\hrule height.#2ex}}}

\def\angleto{\vrule width.035em height2.1ex depth-.56ex\unskip\kern-.6ex\to}
\def\perchc#1{{\raise.4ex\hbox{$\mkern4mu#1{\it\perslsh}_
             {\mkern-5mu\scriptscriptstyle{{\rm o}\!{\rm o}}}^
             {\mkern-12.8mu\scriptscriptstyle{\rm o}}$}}}

\catcode`\@=11 
\def\parenbar{\mathpalette\p@renb@r}
\def\p@renb@r#1#2{\vbox{%
  \ifx#1\scriptscriptstyle \dimen@.7em\dimen@ii.2em\else
  \ifx#1\scriptstyle \dimen@.8em\dimen@ii.25em\else
  \dimen@1em\dimen@ii.4em\fi\fi \offinterlineskip
  \ialign{\hfill##\hfill\cr
    \vbox{\hrule width\dimen@ii}\cr
    \noalign{\vskip-.3ex}%
    \hbox to\dimen@{$\mathchar300\hfil\mathchar301$}\cr
    \noalign{\vskip-.3ex}%
    $#1#2$\cr}}}
\catcode`\@=12 

\newbox\struttbox
\setbox\struttbox=\hbox{\vrule height1.65ex depth.485ex width0pt}
\def\strutt{\relax\ifmmode\copy\struttbox\else\unhcopy\struttbox\fi}
\def\stru#1#2{\relax\ifmmode\hbox{\vrule height#1 depth#2 width0pt}
\else\vrule height#1 depth#2 width0pt\fi}

\def\ronum#1{\uppercase\expandafter{\romannumeral#1}}
\def\ronuml#1{\expandafter{\romannumeral#1}}

\DeclareMathAlphabet{\mathbf}{OT1}{cmr}{bx}{sl}

\renewcommand{\thefootnote}{\arabic{footnote}}
 {\end{list}}
 {\end{list}}
 {\end{list}}

\catcode`\@=11 
\newlength{\@fninsert}
\newlength{\@fnwidth}
\renewcommand{\@makefntext}[1]%
  {\noindent\makebox[\@fninsert][r]{\@makefnmark}\hfil%
  \parbox[t]{\@fnwidth}{#1}}
\catcode`\@=12 
\catcode`\@=11 
\renewcommand\section{\@startsection{section}{1}{\z@}%
                                   {-3.5ex \@plus -1ex \@minus -.2ex}%
                                   {2.3ex \@plus.2ex}%
                                   {\normalfont\Large\bfseries}}
\renewcommand\subsection{\@startsection{subsection}{2}{\z@}%
                                   {-3.25ex\@plus -1ex \@minus -.2ex}%
                                   {1.5ex \@plus .2ex}%
                                   {\normalfont\large\bfseries}}
\renewcommand\subsubsection{\@startsection{subsubsection}{3}{\z@}%
                                   {-3.25ex\@plus -1ex \@minus -.2ex}%
                                   {1.5ex \@plus .2ex}%
                                   {\normalfont\normalsize\bfseries}}
\renewcommand\paragraph{\@startsection{paragraph}{4}{\z@}%
                                   {3.25ex \@plus1ex \@minus.2ex}%
                                   {1.2ex \@plus .2ex}%
                                   {\normalfont\normalsize\bfseries}}
\catcode`\@=12 

\newcommand{\bi}{\begin{itemize}}
\newcommand{\ei}{\end{itemize}}
\newcommand{\be}{\begin{equation}}
\newcommand{\ee}{\end{equation}}
\newcommand{\bea}{\begin{eqnarray}}
\newcommand{\eea}{\end{eqnarray}}

\begin{document}
\setcounter{secnumdepth}{5}
\setcounter{tocdepth}{3}
\selectlanguage{english}
\makeatletter

\pagestyle{empty}
\thispagestyle{empty}
\title{\bf The Intermediate Neutrino Program}
\date{\today}
\clearpage

\renewcommand*{\thefootnote}{\fnsymbol{footnote}}
\footnotetext[1]{Convenor}
\newcommand\ConvenorMark{\footnotemark[1]}
\footnotetext[2]{Organizer}
\newcommand\OrganizerMark{\footnotemark[2]}
\renewcommand*{\thefootnote}{\arabic{footnote}}

\author[72]{C.~Adams}
\author[21]{J.R.~Alonso}
\author[70]{A.M.~Ankowski}
\author[35]{J.A.~Asaadi}
\author[72]{J.~Ashenfelter}
\author[21]{S.N.~Axani}
\author[27]{K.~Babu\protect\ConvenorMark}
\author[5]{C.~Backhouse}
\author[72]{H.R.~Band}
\author[9]{P.S.~Barbeau\protect\ConvenorMark}
% ...
\affil[1]{\mbox{Argonne National Laboratory}}
\affil[2]{\mbox{Bern}}
% ...
\affil[5]{\mbox{California Institute of Technology}}
% ...
\affil[9]{\mbox{Duke University}}
% ...
\affil[21]{\mbox{Massachusetts Institute of Technology}}
% ...
\affil[27]{\mbox{Oklahoma State University}}
% ...
\affil[35]{\mbox{Syracuse University}}
% ...
\affil[70]{\mbox{Virginia Tech}}
% ...
\affil[72]{\mbox{Yale University}}
\maketitle

\clearpage

\parindent 10pt
\pagenumbering{roman}                   
\setcounter{page}{1}
\thispagestyle{plain}
\pagestyle{plain}
\tableofcontents
\clearpage

\pagenumbering{arabic}                   
\setcounter{page}{1}

\section{Introduction}
\label{sec:Introduction}

The US neutrino community gathered at the Workshop on the Intermediate
Neutrino Program (WINP) at Brookhaven National Laboratory February
4--6, 2015 to explore opportunities in neutrino physics over the next
five to ten years. Scientists from particle, astro-particle and
nuclear physics~\footnote{The nuclear physics community is currently
  in the midst of a Long Range Planning process that includes neutrino
  physics.}  participated in the workshop. The US High Energy Physics
community is now launched on the path defined by the 2014 P5 report.
Two of the five P5 Science Drivers motivate neutrino physicists,
\textit{``Pursue the physics associated with neutrino mass''} and
\textit{``Explore the unknown: new particles, interactions, and
  physical principles''}.

Underlying these Drivers is the fact that in spite of tremendous
recent progress in understanding neutrinos, we still lack a complete
picture for the physical behavior and structure of the
neutrino sector. There are many more questions yet to be answered:
What is the neutrino mass ordering? Do neutrinos exhibit the same
matter/antimatter symmetry seen in the charged leptons and quarks
(Dirac fermions) or do they have a completely different structure
(e.g., Majorana fermions)?  Do they violate CP symmetry?  What is the
absolute mass scale of the neutrino sector and is it consistent with
that implied from limits obtained by cosmological observation? Are
experimental hints that there may be additional sterile flavors of
neutrinos valid?  How can we understand neutrino interactions with
nuclei?  Are there non-standard interactions of neutrinos representing
beyond-the-Standard Model physics? 

There are furthermore outstanding questions which can be answered
using neutrinos as a probe, such as: What is the detailed mechanism
behind stellar-collapse supernovae, and what physics can we learn from
a supernova neutrino burst? Is the diffuse flux of neutrinos from past
supernovae consistent with expectations from cosmology?  Where do
ultra high energy neutrinos come from?  What fraction of the Earth's
radiated heat comes from radioactivity?  What fraction of the Sun's
energy comes from the CNO cycle?  What fission isotopes are the main
producers of neutrinos and decay heat in a nuclear reactor core?

In accordance with one specific P5 recommendation, a large international
collaboration (Experimental program at the Long-Baseline Neutrino
Facility, ELBNF) has been formed to perform a long-baseline neutrino
oscillation experiment with an underground liquid argon time
projection chamber and a new neutrino beam from
Fermilab. The main physics goals of ELBNF are to address some of the
outstanding neutrino questions, by measuring the value of
the CP-violating parameter $\delta$ and providing a definitive
determination of the mass hierarchy, as well as to search for baryon
number violation and record a burst of core-collapse supernova
neutrinos. The community is also engaged in development of a short-baseline
neutrino (SBN) program hosted at Fermilab, as a part of the P5
recommendataion for a short baseline neutrino portfolio, to operate
coherently with the long-baseline program in order to address some of
the short-baseline neutrino anomalies and support R\&D towards ELBNF.

The Fermilab long-baseline effort is exciting and compelling, but
is a long-term effort.  Physics results and technical development in
the short term (this decade) are essential for health of the field,
for motivating young scientists and for sustaining innovation.  The
long-baseline oscillation program addresses some of the most critical
questions in particle physics --- yet other physics questions
associated with neutrinos, including the ones listed above, deserve
attention as well.  Some of these are best addressed with smaller,
lower-cost, dedicated experiments which can be completed
on a shorter time scale than ELBNF.  In addition, some issues will
need large, advanced-technology detectors which require
significant R\&D for their realization.

The P5 report states:
\textit{Some of the biggest scientific questions driving the field can only be
addressed by large and mid-scale experiments. However, small-scale
experiments can also address many of the questions related to the
Drivers. These experiments combine timely physics with opportunities
for a broad exposure to new experimental techniques, provide
leadership roles for young scientists, and allow for partnerships
among universities and national laboratories. In our budget exercises,
we main- tained a small projects portfolio to preserve budgetary space
for a number of these important small projects, whose costs are
typically less than \$20M. These projects individually are not large
enough to come under direct P5 review. Small investments in large,
multi-disciplinary projects, as well as early R\&D for some project
concepts, were also accounted for here.}\\
\textit{{Recommendation 4: Maintain a program of projects of all
  scales, from the largest international projects to mid- and
  small-scale projects.}}

WINP explored opportunities for the broad US neutrino community to
pursue physics in the next five to ten years, including projects of
all scales.  In addition to small- and mid-scale project ideas, WINP
considered (consistent with the P5 recommendation) contributions to
large offshore and multi-disciplinary projects, and to R\&D efforts
aimed towards future large experimental concepts.  Some of these ideas
and efforts are synergistic with the ELBNF and SBN efforts; others are
complementary.  Theory efforts related to the proposed long- and
short-term experimental programs were also discussed. WINP also
covered activities such as searches for neutrinoless double beta decay
and direct mass measurements that are part of the neutrino physics
program in nuclear physics.

The workshop was organized into two sets of parallel working group
sessions, divided by physics topics and technology.  Physics working
groups covered topics on Sterile Neutrino, Neutrino Mixing, Neutrino
Interactions, Neutrino Properties and Astrophysical Neutrinos.
Technology sessions were organized into Theory, Short-Baseline
Accelerator Neutrinos, Reactor Neutrinos, Detector R\&D and Source,
Cyclotron and Meson Decay at Rest sessions.  Each working group
formulated a set of bullet points with key findings and
recommendations, which will be summarized in this report.

In addition to members of the particle and nuclear physics
communities, representatives from the agencies participated in the workshop.

A major area of discussion centered around a possible Department of
Energy Office of High Energy Physics Funding Opportunity Announcement
(FOA) relevant to neutrino physics over a five-year timescale.
Discussion sessions included community suggestions on physics topics
and parameters for such a FOA.  The general outcome of this discussion
was that the community believes that a broader rather than a narrower
scientific scope is important for any such FOA.  There was no clear
consensus on the appropriate fraction of available funds to be
allocated to R\&D efforts, but R\&D was felt to be important.
Potential for publishable results --- either physics, or key technical
results that advance the field --- within five years should be the
most important criterion.  There was also community support to enable
proposals for theoretical effort.  These can be highly cost-effective
and may strongly enhance the success of experimental efforts.
The community felt that a suite of new experiments should be
initiated, consistent with the P5 recommendation, to increase the
breadth of the intermediate program beyond SBN at FNAL. A general
sense emerged that some part of this new program should include
sterile neutrino searches complementary to SBN and that part of this
program should include some of the other exciting new initiatives
discussed.

Although much discussion focused on the potential HEP FOA, the broader
neutrino community was actively engaged at the workshop, and it was
clear that there are additional opportunities for activities on an
intermediate time scale.  For example, another area of discussion was
the importance of determining the nature of the neutrino.  Whether
neutrinos are Majorana or Dirac is of fundamental importance to their
mass generation mechanism and to leptogenesis; the search for
neutrinoless double beta decay offers the most promising (and possibly
only plausible) avenue for addressing this question.  The community
recognizes the importance of the Nuclear Science Advisory Committee
Neutrinoless Double Beta Decay subcommittee for providing guidance on a
strategy for implementation of the next generation neutrinoless double
beta decay experiment.  In addition, many other intermediate
activities --- including R\&D, theory, experiments and contributions
to a variety of international efforts --- are supported by the
National Science Foundation and the DOE Office of Nuclear Physics as
well as the Office of High Energy Physics.

\newpage
\section{Physics Working Groups}
\label{sec:Physics WG}

In this section we provide the summary reports of the physics-related
working groups.  These working groups convened in the morning of
February 5, 2015 to discuss the outstanding physics questions in neutrino
physics, with emphasis on those that can be addressed within the next
decade.

\begin{itemize}
  \item Sterile Neutrinos
  \item Three Neutrino Mixing
  \item Neutrino Interactions
  \item Neutrino Properties
  \item Astrophysical Neutrinos
\end{itemize}

\newpage
\subsection{Sterile Neutrinos}
\label{sec:Sterile}

Sterile neutrinos appear in nearly every possible mechanism to explain
neutrino mass. Dark matter and dark energy provide possible evidence
for physics beyond the Standard Model in astrophysics, but neutrino
mass has been the only such evidence so far found in particle
physics. This should make the quest for understanding the origin of
neutrino mass one of the priorities of the field.

Apart from these theoretical considerations, there have been
persistent anomalies, observed in electron neutrino appearance and
electron neutrino disappearance (For the disappearance channel
neutrino and antineutrino oscillation probabilities are equal assuming
CPT invariance and therefore we make no distinction in the remainder of
the document), which in combination can be interpreted as evidence
for one or several sterile neutrinos at around a mass of 1~eV and with
oscillation amplitudes as large as 5--10\%. At the same time, the lack
of observations of muon neutrino disappearance at the relevant L/E
values creates a significant tension in global fits. Currently, no
phenomenological models are known that provide a better fit to the
global data than sterile neutrinos. It therefore seems justified to
use sterile neutrino oscillations as a phenomenological proxy for
considering the required experimental capabilities to resolve these
anomalies. In the event that the existence of sterile neutrinos is not
the source of these anomalies, experiments optimized to eV-scale
oscillations can be expected to also have good sensitivity to probe
other new physics scenarios.

During the Snowmass process, P5 deliberations and WINP discussions,
many different experimental approaches to this problem have been
proposed and some have been studied in quite some detail and some have
reached the prototype stage. Sterile neutrino searches build on
detection techniques developed over the last several years, allowing
for efficient experiment deployment at modest cost in some cases.
Among the many proposed experimental approaches, electron neutrino
disappearance searches using radioactive sources and reactors and some
electron neutrino appearance searches seem to fit well within the
scope of an intermediate neutrino program and would complement the
Fermilab short-baseline program. Thus, a modest investment into
sterile neutrino searches has the potential to discover a fundamental
particle not predicted in the Standard Model for the first time since
the Standard Model was formulated. This would be a paradigm-shifting
discovery opening up vistas on an entirely new continent of
possibilities. The U.S. community is well poised to play a leadership
role and to have a vibrant program in this field.

In the somewhat longer term, and in particular in anticipation of a
potential discovery of sterile neutrinos, directed R\&D towards novel
detector technologies and neutrino sources also would fit the boundary
conditions of an intermediate neutrino program.

The working group's consensus can be summarized in the following five
recommendations:
\begin{enumerate}
\item{Sterile neutrinos are well motivated in many extensions of the
  Standard Model.  Persistent experimental anomalies have focused
  attention on the eV mass scale.  This makes sterile neutrinos the
  subject of low-risk but potentially high-reward experiments.
  Therefore, the P5 Planning Report recommends a targeted set of
  short-term, small-scale experiments.}
\item{Direct tests of existing anomalies should seek to demonstrate
  the sterile neutrino's oscillatory nature via signatures in energy
  and/or baseline.}
\item{Experiments designed to test both the $\nu_{\mu}$ to $\nu_e$
  appearance and $\nu_e$ disappearance channels are needed.  We must
  ensure that any pion decay beam program has optimized $\nu_{\mu}$
  disappearance sensitivity.}
\item{Below the \$5M level, individual experiments can have an impact on
  oscillation physics results within the WINP time constraints.  In
  the \$5--10M range, several proposed efforts have the potential to
  provide extensive coverage of the suggested oscillation parameter
  space.}
\item{Short-term investment in detector and source R\&D towards future
  sterile oscillation experiments could reduce risk, lead to long-term
  cost savings, and provide the foundation for precision measurements
  in the case of observation of sterile neutrino oscillations.}
\end{enumerate}

\newpage
\subsection{Three Neutrino Mixing}
\label{sec:3NuMixing}

Three neutrino mixing has been well established by a variety of
experiments yielding consistent results. Besides the overall neutrino
mass scale, there are three parameters that are still unknown in this
framework and which can be better addressed in neutrino oscillation
experiments: the neutrino mass ordering, $\delta_{CP}$, and the octant
of $\theta_{23}$.  The determination of these unknowns is of key
importance to understanding the mechanism responsible for neutrino
mass and potentially the generation of the matter-antimatter
asymmetry in the early universe.  The suite of presently running
experiments are either directly exploring the three unknowns, or are
producing results that will contribute to these measurements
by reducing systematic and theoretical uncertainties (neutrino
interaction measurements for example), or by precisely measuring
related oscillation parameters ($\theta_{13}$, $\theta_{23}$, $|\Delta
m^{2}_{\textnormal{atm}}|$).  Proposed and running experiments can
potentially measure the ordering and $\theta_{23}$ octant in the next
decade.  The measurement of $\delta_{CP}$ is likely to take longer and
require a dedicated large-scale experiment.  If discrepancies are
found in the measurement of one or more of these fundamental
parameters across different experiments, it could point to new
physics.

Recent results from the Daya Bay Reactor Neutrino
Experiment~\cite{Zhang:2015fya} include the most precise measurement
of $\sin^{2}2\theta_{13}$ to date and a measurement of $|\Delta
m^{2}_{\textnormal{atm}}|$ that is comparable in precision and
consistent with the long-baseline measurements.  By Daya Bay's
expected end date in 2017, it will provide the most precise
measurement of $\sin^{2}2\theta_{13}$ for many years to come.  The
Double Chooz experiment has recently installed their near detector,
which will greatly increase the precision of their
$\sin^22\theta_{13}$ measurements~\cite{Abe:2014bwa}. Double
Chooz has the unique ability to take reactor-off data due to having
only two reactor cores, allowing a strong constraint on the background
in their $\theta_{13}$ analysis.

The combined data from the MINOS long-baseline experiment using the
low-energy NuMI beam and atmospheric neutrino data in the MINOS far
detector provides the most precise measurement of the atmospheric mass
splitting~\cite{Adamson:2014vgd}.  MINOS+ is currently running in the
medium-energy NuMI beam.  Including these data in the fit will further
improve the precision.  Furthermore, the high statistics in the energy
region just above the oscillation maximum puts MINOS+ in a unique
position to test the validity of the three neutrino mixing model. T2K
can also measure the atmospheric parameters and has made the most
precise measurement of $\sin^2\theta_{23}$.  T2K also made the first
observation of electron neutrino appearance.  Assuming the value of
$\theta_{13}$ from reactor data, T2K can exclude some values of
$\delta_{CP}$ at the 90\% confidence level~\cite{Abe:2015awa}.  The
NOvA experiment at Fermilab has only just started running, but has the
potential to determine the octant or the mass ordering within some
range of the parameters in particular in combination with data from
both T2K and MINOS+.

Super-Kamiokande continues to use atmospheric neutrino data to address
the three main questions in three neutrino mixing and has recently
included combined fits with T2K data~\cite{Wendell:2014dka}.  A
proposed upgrade for Super-Kamiokande in the intermediate program would involve
adding gadolinium to the water to identify electron antineutrinos via
the inverse beta decay interaction~\cite{Beacom:2003nk} enabling
detection of diffuse supernova neutrinos. IceCube has recently
presented results on the atmospheric mixing parameters that are
consistent with and comparable in precision with results from
Super-Kamiokande~\cite{Aartsen:2014yll}.

It is important to realize that precision measurements of oscillation
parameters made by currently running experiments can have a
significant effect on future projects --- either in achievable
precision, detector or beam design or in the running conditions (for
example, neutrino beam vs antineutrino beam).

Besides these oscillation experiments directly addressing the
determination of neutrino parameters, experiments collecting neutrino
interaction data are also of great importance for constraining the models
that are used in oscillation experiments for true-to-visible energy
conversions, predictions of signal and background rates, etc.  MINERvA
is a dedicated neutrino interaction experiment that provides
constraints for oscillation experiments directly and for neutrino
event generators used in neutrino simulations.  The MINERvA
collaboration is currently finishing the analysis of their low-energy
NuMI data (for example,~\cite{Walton:2014esl}) while taking data in
the NuMI medium-energy beam.  A potential upgrade to the MINERvA
detector is being proposed for the intermediate program.  CAPTAIN, a
liquid argon TPC~\cite{Berns:2013usa} (LArTPC), would be combined with the
MINERvA detector to measure neutrino-argon interactions in an energy
range relevant for oscillation physics in ELBNF.

The US-NA61 program is a funded proposal to collaborate with the
NA61/SHINE experiment at CERN~\cite{Gazdzicki:2014bxa} to make hadron
production measurements important for the US neutrino program.  They
will expose targets and replicas of targets used at Fermilab to the
NA61 hadron beam.  Measurements of pion (and other hadron) spectra in
p+C interactions can be used to tune models of primary interactions in
the target and reduce uncertainties on the initial flux in oscillation
experiments.

The NuPRISM collaboration is proposing an experimental method to
remove neutrino interaction uncertainties from oscillation
experiments using water Cherenkov detectors~\cite{Bhadra:2014oma}.  NuPRISM would measure neutrino
interactions over a continuous range of off-axis angles and use these
measurements to provide a direct measurement of the far detector
lepton kinematics for any given set of oscillation parameters.  With
this, they can mostly remove neutrino interaction modeling
uncertainties from oscillation measurements.  This project is being
proposed for the intermediate program.

There is also a group of longer term planned experiments designed to
measure the last unknown elements of the three neutrino mixing model.
These include JUNO, Hyper-K, PINGU, \textsc{Theia}, Daedalus, and ELBNF. Of these experiments, JUNO is the only one currently under construction.  
R\&D for these large projects will be an important part of the intermediate
neutrino program.

Within this context, the most relevant issues for these experiments
to seek funding in this intermediate time scale are:
\vspace*{-0.4cm}
\begin{itemize}
\item Discovery potential and/or improved physics reach over other experiments
\vspace*{-0.4cm}
\item Leverage existing resources \vspace*{-0.4cm}
\item Visible and significant US participation \vspace*{-0.4cm}
\item Low technical risk/familiar technology \vspace*{-0.4cm}
\item Potential for each to result in several graduate student theses \vspace*{-0.4cm}
\item US contribution of approximately \$1M each \vspace*{-0.4cm}
\end{itemize}

Of the experiments discussed in the three neutrino mixing working group, there are only a few proposals that will seek funding for the intermediate program.  The above requirements are well fulfilled by these proposals, including proposed experiments
measuring cross sections such as NuPrism and CAPTAIN-MINERvA, as well
as upgrades to existing detectors like gadolinium in Super-Kamiokande.  In
addition, R\&D for future projects described above is
relevant, in particular for those which can lead to clearly definable US
roles in international collaborations.

\newpage
\subsection{Neutrino Interactions}
\label{sec:Interactions}

Significant investment is being made using
accelerator-based sources of neutrinos to critically test the current
3-neutrino mixing picture and determine whether or not neutrinos
violate CP.
The detectors associated with this program use heavy nuclei to achieve
the required event rates and modeling the interaction of neutrinos
with these heavy nuclei is required.  Now that neutrino
oscillation experiments are evolving from the discovery to the
precision stage, understanding the challenging role of the nucleus in
neutrino interactions has become an imperative. It is needed to sort
signal from background and to identify and minimize systematic
uncertainties in neutrino oscillation investigations.
It has therefore become essential to establish a more robust program
of characterizing neutrino-nucleus interactions and promptly
integrating these results into event generators. This difficult task
requires the cooperative effort of theory and experiment from both
nuclear and high-energy physics.

Nuclear theorists are providing far more realistic nuclear models than
the relativistic Fermi gas, which has been the standard in event
generators, however, many aspects of the theory are still not at the
required level.  Event generators have begun the process of including
these improved although still incomplete advances, but manpower
problems are a hindrance. Thus, agreement of generator predictions
with data still remains a challenge and there is real concern with the
accuracy of neutrino energy
estimations~\cite{Martini:2011wp,Lalakulich:2012hs,Shneor:2007tu}.
This increased uncertainty in assigning the incident neutrino energy
will become a serious issue for oscillation measurements in the future
as statistics increase and the need for greater precision is
required~\cite{1412.4294}.  Furthermore, the HEP oscillation program
requires current neutrino nucleus theoretical developments to be
extended to higher-A such as Ar, to more relativistic energies and
momenta and to include the yield of resonance production cross
sections above the delta.  Significantly, this important work by
nuclear theorists has to be packaged in a form that can be swiftly
incorporated into neutrino event generators. Although it is not yet
clear how to do this, extensive collaboration between nuclear
theorists, builders of event generators and neutrino
experimentalists will be required. To achieve this, communication
between the NP and HEP communities should be improved so that new
theoretical advances can be more readily adopted.

In addition, the theory discussed above only describes inclusive cross
sections. Calculating the wide variety of exclusive final states
exceeds any current capability. Dealing with final state interactions
(FSI) will require extensive recourse to phenomenology.  Significantly
more complete higher energy neutrino cross section data will be
needed, especially for argon targets such as the proposed
CAPTAIN-MINERvA collaboration.  Additional input will likely come from
electron scattering carried out at kinematics similar to those
encountered in neutrino experiments. The electron beam at Jefferson
Laboratory is ideal and there is a recent JLAB
proposal~\cite{Benhar:2014nca} to carry out (e,e$^\prime$p)
measurements on $^{40}Ar$. The event generators must then incorporate
these data into appropriate models for existing and planned
experiments.
It is critical to benchmark generators against both accelerator-based
neutrino-nucleus interaction measurements and electron-nucleus
interaction measurements.
The current experimental neutrino interaction program 
continues to provide important data and should be supported to its
conclusion.  To further refine the nuclear model in event generators,
future neutrino interaction measurements that span a range of neutrino
and antineutrino beam energies as well as target materials will be
needed to bring the current knowledge to the level required to reach
the goals of the long baseline oscillation program. The progress in
developing a broad international GENIE collaboration with a core group
at FNAL is encouraging.
Understanding the subtleties of the nuclear environment and their
effect on neutrino experiments needs the input of nuclear physics
theorists specializing in this topic. It is important to create an
established procedure that allows nuclear theorists to join neutrino
generator experts and neutrino experimentalists in working toward this
goal. NuSTEC (Neutrino Scattering Theory Experiment Collaboration,
http://nustec2014.phys.vt.edu) has been established directly to
provide this environment.
A current example of such a project is the work on Green's Function
Monte Carlo techniques by a collaboration of Argonne, Jefferson Lab,
and Los Alamos nuclear theorists with Fermilab neutrino
experimentalists.  A white paper (Lovato, A. et. al. "Quantum Monte
Carlo Methods for Neutrino-nucleus Interactions",
FERMILAB-FN-0997-ND-T, JLAB-THY-15-2012, LA-UR-15-21054) from a
collaboration of NP theorists and HEP neutrino experimentalists has
been submitted to DOE-HEP and DOE-NP and is now being expanded to a
full joint HEP-NP proposal to be submitted to the DOE.

A summary of this Neutrino Interactions working group's discussions in
the form of a prioritized list follows:
\begin{enumerate}
\item The Neutrino-Nucleus Interaction is the least understood
  component of a detector’s response to neutrinos.

\item Improvements of nuclear models by nuclear theorists are
  essential. This can most efficiently be accomplished with additional
  financial support of NP theorists working in this area to provide a
  more robust model to meet the requirements of the oscillation
  program.  Rapidly incorporating these improvements in event
  generators is equally important and requires a collaborative effort
  of the HEP and NP communities.

\item The current experimental neutrino interaction program (MINERvA,
  NOvA-ND, MicroBooNE, T2K Near Detector) continues to provide
  important data and should be supported to its conclusion.  This
  includes efforts to improve the precision with which the neutrino
  flux is known.

\item The critical role of neutrino nucleus event generators needs to
  be emphasized and more community resources devoted to keeping them
  widely available, accurate, transparent, and current.

\item Future neutrino interaction measurements such as the Fermilab
  short-baseline program (SBN) and the CAPTAIN-MINERvA experiment are
  needed to extend the current program of GeV-scale neutrino
  interactions.  The feasibility of a high-statistics deuterium
  experiment should be considered.  Current and future
  long-and-short-baseline neutrino oscillation programs should
  evaluate and articulate what additional neutrino nucleus interaction
  data is required to meet their ambitious goals and support
  experiments that provide this data.

\item Measurements and theoretical work are also needed to
  characterize neutrino interactions in the low energy regime ($ \leq
  100 $ MeV). This regime is especially relevant for core-collapse
  supernova neutrinos, and understanding is essential for development
  of future underground detectors.  This is also an area of
  collaboration where NP would bring in critical expertise.

\end{enumerate}

\newpage
\subsection{Neutrino Properties}
\label{sec:Properties}

Neutrinos are the most enigmatic particles in the Standard
Model. Identifying their properties may go beyond merely filling out
the few missing entries in the Particle Data Book. The absolute value
of the neutrino mass has cosmological implications. The magnetic
moment of the neutrino is a sensitive probe of TeV-scale physics
beyond the Standard Model. And identifying the quantum field nature of
the neutrino, i.e. whether it is a Dirac or Majorana fermion, may
determine if Lepton Number is a fundamental symmetry of Nature, and
shed light on the long-standing puzzle of the abundance of matter in
the visible Universe.

\subsubsection{Absolute neutrino mass}

Absolute values of the neutrino masses are among the last unknown
parameters of the new ($\nu$) Standard Model.  The combined results of
absolute neutrino mass searches, neutrinoless double beta decay
searches, and cosmology provide an extraordinary constraint on the
neutrino mass spectrum and models of the neutrino
mass~\cite{Bahcall:2004ip}.  Several experiments have been discussed
at the workshop: KATRIN, Project-8 and electron capture in
$^{163}$Ho. In the US, direct measurements of the neutrino mass are
generally supported by DOE-NP and NSF.

KATRIN represents the state of the art of the presently available
technology. It will reach its ultimate sensitivity ($0.2$~eV at 90\%
C.L., $0.35$~eV for a 5$\sigma$ discovery) by the beginning of the
next decade. This sensitivity is comparable to precision available
from cosmological and neutrinoless double-beta decay constraints
expected on a similar timescale, which may provide even deeper
understanding of neutrinos. KATRIN is currently in commissioning, and is expected
to start operations with the tritium source in 2016.

Techniques that may ultimately exceed KATRIN precision are in
development. Project-8 is a novel idea of measuring the momentum
spectrum near the tritium end-point using microwave cyclotron
radiation spectroscopy~\cite{Monreal:2009za}. The proof-of-principle
demonstration with a $^{83m}$Kr source has recently been
reported~\cite{Asner:2014cwa} in a small-volume trap. First operations
with a tritium source are planned and sensitivity similar to KATRIN is
possible by the end of the decade. Ultimately, a large
$\mathcal{O}(5~\mathrm{m}^3)$ volume experiment with an atomic tritium
source may start approaching the sensitivity to the neutrino mass
below the inverted hierarchy region. On the timescale of next decade,
such measurement would be complementary to direct measurements of the
hierarchy at ELBNF and cosmological constraints on the neutrino mass
of similar sensitivity.

Experiments looking to measure the end-point of the electron capture
spectrum, e.g. in $^{163}$Ho using bolometric micro-calorimeters are
in development (ECHO and HOLMES in Europe, NuMECS in the US). These
experiments employ large arrays of low-mass bolometers with a
$^{163}$Ho source embedded in the absorber. Novel readout techniques
are being explored to enable multiplexed readout of the a large number
of channels; these efforts can benefit from synergy with the
large-scale CMB arrays (supported by DOE-HEP through the CMB-S4
experiments). Large-scale production of $^{163}$Ho sources at reactor
facilities needs further development and would benefit from the
domestic isotope production program, e.g. within the scope of DOE-NP
Isotope Program (Isotope Development \& Production for Research and
Applications or IDPRA).  Micro-calorimeter experiments aim to scale to
10k--100k channels in about a decade, promising to reach sensitivity
to the neutrino mass below $0.1$~eV.

\subsubsection{Coherent elastic neutrino-nucleus scattering (CEvNS)
  and neutrino magnetic moment}

Coherent Elastic Neutrino-Nucleus Scattering (CEvNS) is a yet-unobserved
process that is cleanly predicted by the Standard Model.  The US
community is leading the effort towards the first observation of this
process, which may be possible with relatively modest investments in
the next few years. Coherent scattering experiments are also
sensitive to an anomalous magnetic moment of the neutrino. Detection
requires low energy thresholds and high-intensity neutrino sources,
thus there is complementarity with several other programs (neutrino
scattering, dark matter detection). Once observed, CEvNS could help
constrain models with non-standard neutrino interactions, have
sensitivity to nuclear weak charge, and would offer complementary
constraints on the sterile neutrino sector. 

A variety of source/detector configurations are being considered in the
search for CEvNS. Measurements with low-energy sources would rely
on the low-threshold detectors developed for dark matter searches at
reactors (e.g. RICOCHET) or with high-intensity radioactive
sources (e.g. $^{51}$Cr deployed in LZ). Measurements at higher
energies at accelerators (e.g. COHERENT at SNS and CENNS at Fermilab)
would use more conventional technologies. Searches for CEvNS match
well to the parameters of the program discussed at this workshop: they
can result in the first observation and sensitive limits on the
neutrino magnetic moment within 5--10 year timeframe, and with modest
investments ($<$\$10M). 

The RICOCHET experiment leverages significant R\&D and engineering on
low energy threshold detectors for SuperCDMS. It would deploy a tower
of six SuperCDMS detectors near a research reactor. With energy
thresholds as low as 100~eV$_{nr}$ already achieved at SuperCDMS,
several thousand events per month could be observed, depending on the
reactor and distance to the core. A critical issue is understanding
the in-situ backgrounds and could benefit from synergy with
short-baseline searches for sterile neutrinos at reactors.  Proponents
expect first deployment in the next few years and results on a 5--10
year timescale.  RICOCHET Phase-II~\cite{Formaggio:2011jt} would
operate underground with an intense electron capture source
(e.g. $^{51}$Cr), which requires development of even lower threshold
detectors and large active mass.

Another possibility for deployment of an intense electron capture
source such as $^{51}$Cr is the veto region of the LZ dark matter
detector~\cite{Coloma:2014hka}. With a 5~MCi source 1--2m from the
active volume and the low LZ energy threshold, detection
of CEvNS is possible with high significance. The experiment could
also place limits on the neutrino magnetic moment in the range
of (3--4)$\times10^{-12}\mu_B$, better than any terrestrial limit to
date and comparable in sensitivity to astrophysical constraints. This
effort is complementary with the sterile neutrino program and
could benefit from an enhanced US isotope production and
isotope enrichment program.

Experiments with intense accelerator-based sources, such as the
COHERENT~\cite{Akimov:2013yow} experiment with a DAR source at SNS and
CENNS~\cite{Brice:2013fwa} at the BNB facility at Fermilab offers
perhaps the most expeditious way of detecting CEvNS. COHERENT
would deploy multiple neutrino detectors in the vicinity of an intense
neutrino source from the SNS. Ge detectors (a test module from
the {\sc Majorana Demonstrator\/} neutrinoless double-beta decay
experiment), CsI crystals, as well as a 100~kg LXe 2-phase TPC are
being considered. The collaboration has identified several possible
deployment sites at SNS and has measured in-situ backgrounds.  The
Phase-I program is  ongoing, with the first neutrino scattering
results expected this year, and a possibility to discover CEvNS
within three years. The collaboration is receiving generous support
from ORNL and commitment from other national labs, as well as in-kind
contributions from international partners.

\subsubsection{Majorana/Dirac nature of neutrinos (Neutrinoless
  Double-Beta Decay)}

Searches for Neutrinoless Double-Beta Decay ($0\nu\beta\beta$) aim to
discover whether Lepton Number is a fundamental symmetry of nature or
is violated, and to determine the Dirac or Majorana nature of
neutrinos. The current generation of experiments will search for
$0\nu\beta\beta$ with a sensitivity to the effective Majorana mass of
order 100 meV~\cite{deGouvea:2013onf}. The next generation of
experiments will aim for an order of magnitude improvement in
sensitivity to the effective Majorana mass. With ``tonne-scale''
isotopic mass and ultra-low background next-generation experiments can
discover $0\nu\beta\beta$ if it proceeds via light Majorana neutrino
exchange and if the lightest neutrino mass is above $\sim50$ meV, or
if the spectrum of neutrino masses is
``inverted''~\cite{Bahcall:2004ip}.  Even if neither of these two
conditions is respected in Nature, a discovery is possible if other
mechanisms contribute to the decay. While the potential for discovery
of the Lepton Number Violation in $0\nu\beta\beta$ is independent of
the neutrino mass hierarchy, sensitivity to the smallest possible
masses consistent with the inverted hierarchy is a milestone goal:
with cosmological and terrestrial constraints on the neutrino masses
and the mass hierarchy expected next decade, the next-generation
$0\nu\beta\beta$ experiments can be definitive.

Planning for the next-generation tonne-scale $0\nu\beta\beta$
experiment now is timely, and would help maintain US leadership in the
race for a Nobel-caliber discovery. The US community is gearing
towards selecting the leading candidates for one or more
next-generation experiments, under the stewardship of DOE-NP.  A
targeted program of R\&D activities towards mature concepts of
next-generation experiments is one of the priorities identified by the
recently convened NSAC-NLDBD committee.

Experimental efforts with a significant US participation were
presented at this workshop: CUORE and CUPID, \textsc{Majorana
  Demonstrator\/} and 1 tonne Ge effort, EXO-200 and nEXO, NEXT,
KamLAND-Zen and NuDot, SNO+ and \textsc{Theia}, SuperNEMO. The talks
focused on the current status and on the R\&D needed for the
next-generation, tonne-scale experiments.

The R\&D effort towards definition of the next-generation
$0\nu\beta\beta$ experiment is proceeding at a vigorous pace. There is
possible synergy with the R\&D activities in other areas of neutrino
science as well as broader detector technology efforts within DOE-NP
and DOE-HEP. For instance, there is a general need for reliable
designs of high-voltage distribution systems in noble-gas liquid and
gas TPCs, which could benefit from cooperation between
$0\nu\beta\beta$, dark matter and ELBNF experiments. Development of
large liquid scintillator and WbLS detectors, and in particular
techniques for isotope loading, improvements in light yield and
transparency are common to KamLAND-Zen, SNO+, \textsc{Theia}, NuDot
and others. Low-background techniques, including low-mass, low-noise
custom electronics which could be deployed near sensitive detector
volumes could benefit both $0\nu\beta\beta$  and dark matter
efforts. Those efforts also benefit from ongoing cooperation in
improving radio-assay capabilities and sensitivities. Novel sensor
technology development cuts across disciplines, from neutrino science
to large-scale CMB experiments, to $0\nu\beta\beta$ and dark matter
experiments. Finally, availability of a large quantity of isotopically
enrichment material will be critical for most future $0\nu\beta\beta$
experiments. Further development of domestic isotopic enrichment
capabilities, building on the existing DOE-NP Isotope Program (IDPRA),
would be beneficial to the broad US-based neutrino physics program.

\newpage
\subsection{Astrophysical Neutrinos}
\label{sec:Astrophysical}

\subsubsection{Low Energy Astrophysical Neutrinos}
\label{sec:Astrophysical_low}

Low energy astrophysical neutrinos (typically below about 100 MeV)
result from stars either: a) going about their usual business of
nuclear fusion and emitting a steady stream of solar neutrinos, or b)
living fast, dying young, and leaving a good-looking corpse, in the
process producing a spectacular burst of supernova neutrinos. These
astrophysical neutrinos are important physics messengers, carrying
information on extreme environments and complex processes that would
be otherwise entirely inaccessible, namely what is going on in the
center of stars as they live and die. They have a storied history in
particle physics: the first observations of solar neutrinos (in the
1960's) and supernova neutrinos (in the 1980's) resulted in a shared
2002 Nobel Prize in physics. Famously, the long-standing ``Solar
Neutrino Problem'' turned out to be the first indication of physics
beyond the standard model, a glimpse at the effect of neutrino
oscillations.

What remains to be done on the intermediate time (and money) scale?
For solar neutrinos the main job for the next few years will be to
measure neutrinos produced by the CNO cycle and to explore the MSW
resonance in the sun. For supernova neutrinos from an explosion in our
galaxy, the most important thing is to be ready to collect as much of
this precious data as possible when it arrives, preferably with
complementary technologies to study the various neutrino flavors
and features of the event, particularly the initial neutronization
burst and the final collapse to a black hole. While waiting for the
next galactic burst to arrive the diffuse flux of supernova neutrinos
from ancient supernova explosions can also be collected, provided a
detector is big enough and sensitive enough.

In the coming five years it is expected that an upgraded
Super-Kamiokande enhanced with gadolinium, Borexino, and SNO+ will be
the major solar neutrino experiments in operation; over the same
period Super-Kamiokande with gadolinium, SNO+, and WATCHMAN will be
the main ``new'' projects with significant supernova neutrino
sensitivity. At the same time, CAPTAIN will be providing insight into
the supernova neutrino capabilities of liquid argon detectors like
ELBNF, while R\&D on water-based liquid scintillator should prove a
good investment for longer-term projects like \textsc{Theia}, which would be
capable of studying both solar and supernova neutrinos.

\subsubsection{High Energy Astrophysical Neutrinos}
\label{sec:Astrophysical_high}

High energy astrophysical neutrinos ($\sim$100~MeV--$10^{20}$~eV) can be
separated into two categories: 
\begin{enumerate}
  \item The ``atmospheric neutrinos'', which are neutrinos produced terrestrially in the earth's
atmosphere as a result of high energy cosmic rays interacting with the
atmosphere, and have been detected with energies of $\sim$100~MeV to over
100~TeV. 
   \item The ``high energy astrophysical neutrinos'', which are
neutrinos produced and originating from extra terrestrial sources, for
example from high energy astrophysical processes in distant
astrophysical objects, high energy cosmic ray interactions in the
universe or neutrino production by exotic matter and particle physics
processes such as dark matter annihilation. The neutrino energies of
these processes are observable above backgrounds at energies of
$\sim$10~TeV and above, and have recently been observed for the first
time at energies up to 2~PeV. This newly discovered neutrino source
holds promise for determining the origin of the highest energy cosmic
rays that has remained a mystery for over 100 years and to open a new
energy window for observing astrophysical objects and the physical
processes responsible for producing the highest energy particles in
the universe.
\end{enumerate}

During this workshop the working group identified three areas
appropriate for possible funding by a program aimed at the
``intermediate'' time and money scale. These include: 
\begin{enumerate}
  \item Funding for a cosmogenic (GZK) neutrino detector using radio detection
techniques. These projects include the ARA and ARRIANA programs both
located in Antarctica.
  \item Funding for R\&D on photodetector,
instrumentation and deployment systems development for large channel
count and physically large volume detectors. Possible projects include
IceCube-Gen2 at South Pole and for related technologies in other
science categories such as PINGU, CHIPS, \textsc{Theia} and others. 
  \item Funding for theoretical work on neutrino production at high energies in
atmospheric neutrinos, and in particular the uncertain flux of the so
called ``prompt'' neutrinos from charm production processes in
cosmic ray interactions with the atmosphere.
\end{enumerate}

\newpage
\section{Technology Working Groups}
\label{sec:TechnologyWG}

In this section we provide the summary reports of the
technology-related working groups which convened in afternoon
parallel sessions on Febrary 5, 2015. These working groups covered
methods, including theory and R\&D, for addressing the physics goals
discussed in the physics working groups.

\begin{itemize}
  \item Short-Baseline Accelerator Neutrinos
  \item Reactor Neutrinos
  \item Source, Cyclotron and Meson Decay at Rest Neutrinos
  \item Neutrino Detector R\&D
  \item Neutrino Theory
\end{itemize}

\newpage
\subsection{Short-Baseline Accelerator Neutrinos}
\label{sec:AcceleratorSB}

Accelerator decay-in-flight (DIF) neutrino beams provide an excellent
opportunity for near-term, cost-effective neutrino experiments
pursuing a variety of exciting physics and detector development goals.
The US is home to two existing neutrino beams, both located at
Fermilab.  The Booster neutrino beam (BNB) has the significant
advantage of being at shallow depth and parallel to the ground,
enabling the deployment of multiple detectors at different baselines
for relatively modest construction costs.  Improvements to the
performance of the BNB are currently under consideration and would
significantly strengthen all experiments that utilize this beam.
Preliminary studies indicate that a significant flux increase is
feasible by fitting a second horn in the existing beam facility,
although a detailed schedule and cost estimate needs to be prepared as
recently requested by the Fermilab Physics Advisory Committee
(PAC). New experiments on the Main Injector neutrino beam (NuMI) are
also possible, but are limited by existing space in the underground
near detector cavern.

A major element of the intermediate neutrino program will be the
Short-Baseline Neutrino (SBN) program of three liquid argon detectors
along the BNB~\cite{SBNProposal}.  The SBN program can resolve a class
of experimental anomalies in neutrino physics and perform the most
sensitive searches to date for sterile neutrinos at the eV mass-scale
through appearance and disappearance oscillation channels in an
accelerator DIF neutrino beam.  Additional physics includes the study
of neutrino-argon cross sections with millions of interactions using
the neutrino fluxes of the BNB (on-axis) and NuMI (off-axis) beams.
This program brings together an international team of scientists and
engineers to advance LArTPC technology for neutrino physics while
utilizing its capabilities to explore one of the exciting open
questions in neutrino physics today.  The SBN program was awarded
Stage-1 approval at Fermilab in February 2015.  Preparation of the
three detectors is proceeding rapidly in order to match a tight time
schedule that anticipates first data in 2018.  MicroBooNE is now
installed on the BNB and ready for commissioning, ICARUS is being
refurbished at CERN in preparation for operations at shallow depth,
and detailed designs are being developed for the near detector,
LAr1-ND.

The Accelerator Neutrino Neutron Interaction Experiment
(ANNIE)~\cite{Anghel:2014ynd} is another proposal to utilize the
on-axis flux of the Booster neutrino beam.  ANNIE is a gadolinium
doped water detector instrumented with advanced photodetectors that
will measure neutron production in GeV-scale neutrino interactions,
something presently not well understood and that represents a limiting
factor in proton decay and supernova neutrino measurements in large
water detectors.  ANNIE also presents an opportunity to demonstrate
new Large Area Picosecond Photodetector (LAPPD)
technology~\cite{lappd}, now being commercialized through the DOE STTR
program, in the context of a neutrino detector. ANNIE was awarded
Stage-1 approval following the January meeting of the Fermilab PAC.


The CAPTAIN detector~\cite{Berns:2013usa}, a 5 ton LArTPC being built
at Los Alamos National Laboratory, anticipates running in both the BNB
and NuMI neutrino beams.  By positioning the detector far off-axis at
the BNB, CAPTAIN can study low energy neutrino interactions (10s of
MeV) to better understand neutrino interactions in the energy range of
supernova neutrinos that may be observed in future large LAr
detectors.  The CAPTAIN-MINERvA experiment involves positioning the
CAPTAIN detector in front of the existing MINERvA detector in the NuMI
near detector hall to make precision measurements of neutrino-argon
cross sections in the multi-GeV energy range.


The NuPRISM detector~\cite{Bhadra:2014oma} represents a novel approach
to substantially reducing the impact of neutrino cross section model
uncertainties in long-baseline oscillation experiments, and has been
proposed as part of the near detector complex at J-PARC in Japan. By
sampling the neutrino beam at angles from 1--4 degrees off-axis,
representing different ranges of neutrino energy, NuPRISM can map out
the relationship between neutrino energy and the observed lepton
kinematics. NuPRISM, in combination with the existing T2K near
detector, also has sensitivity to sterile neutrino oscillations by
sampling the beam at the same $L$ but different $E_{\nu}$
ranges. Finally, NuPRISM can construct mono-energetic beams to make
unique neutrino cross section measurements, such as the first ever
neutral current cross sections as a function of neutrino energy, and
the separation of traditional CCQE events from interactions involving
several nucleons.

Accelerator neutrino beam facilities and precision neutrino detectors
present another exciting opportunity for the interim physics program.
By running in a `beam-dump' mode (steering the proton beam off the
target and into the beamline absorber), searches can be performed for
dark sector particle production with reduced neutrino backgrounds.
The MiniBooNE experiment has recently collected data in this mode and
is currently analyzing the
results~\cite{Dharmapalan:2012xp,Thornton:2014ufa}.  The LAr detectors
of the SBN program present an opportunity to perform similar searches
in the future with enhanced sensitivity.

The need to understand the physics being pursued with these
short-baseline accelerator beam experiments (anomalies, sterile
neutrinos, neutron production, neutrino-nucleus cross sections) and
the importance of developing detector technologies for the future
neutrino program motivate an aggressive time scale for each of these
experiments, making them ideal for an interim neutrino program on the
road to the next-generation US long-baseline experiment.

\newpage
\pagebreak
\subsection{Reactor Neutrinos}
\label{sec:Reactor}

Reactor neutrinos have played a central role in the history and our
understanding of neutrinos.  From the first experimental observation
of the neutrino by Reines and Cowan at the Savannah reactor in the
1950's~\cite{Reines:1960pr} to the demonstration of neutrino
oscillations with KamLAND~\cite{Abe:2008aa} and the recent precision
measurement of the neutrino mixing angle
$\theta_{13}$~\cite{Abe:2012tg, An:2013zwz}, reactor neutrinos have
provided us with a unique tool for discovery for over five
decades. Reactor experiments have advanced our understanding of the
3-neutrino framework and enabled the most precise measurement of
neutrino mixing.

Emission of electron antineutrinos from reactors provides a
flavor-pure source of antineutrinos with energies up to
$\sim$8~MeV. Reactor neutrinos are studied at distances as close as
several meters from the reactor core and up to hundreds of kilometers.
The abundant flux of antineutrinos from a nuclear reactor allows
experiments to probe for new physics such as sterile neutrinos and
neutrino magnetic moments, observe coherent neutrino scattering, and
determine the neutrino mass hierarchy. The detection and study of
reactor antineutrinos allows the monitoring and study of the power and
fuel composition of reactors and finds application in
non-proliferation and safeguards.

\subsubsection{Physics Reach and Discovery Potential}
Over the next decade neutrino experiments will make precision tests of
3-neutrino oscillations, aim to understand anomalous signatures in the
current suite of neutrino data, determine the mass hierarchy and
search for new CP violation. Reactor experiments will play a unique
role in this program and provide discovery potential of new physics at
modest costs. The observation of sterile neutrinos or other new
physics would be a paradigm shift and set the course of particle
physics for years to come.  The resolution of the neutrino mass
hierarchy is fundamental to understanding the neutrino mass spectrum
and will provide important input to the search for neutrinoless double
decay. Precision measurements of neutrino mixing parameters and
knowledge of the mass hierarchy will help maximize the physics reach
of long-baseline experiments and are fundamental to any extensions to
the Standard Model.

Measurements of the reactor neutrino flux and spectrum compared to
recent models of reactor antineutrino production have revealed
discrepancies in both the total measured reactor antineutrino flux as
well as the energy spectrum of neutrinos. The observed flux is found
to be $\sim$5--6\% low~\cite{Mention:2011rk, Zhang:2013ela} while the
recent $\theta_{13}$ experiments have revealed a distortion in the
4--6~MeV region of the spectrum. The observed discrepancies may point
to new physics such as eV-scale sterile neutrinos. The observed
spectral features already indicate that nuclear models in the
predictions of the antineutrino flux from reactors are incomplete. The
latter is also supported by recent measurements of the antineutrino
spectrum from selected nuclei~\cite{MTAS1,MTAS2,MTAS3}. A
well-controlled measurement of the antineutrino flux and spectrum at
short-baselines can directly address these questions simultaneously
with a unique measurement of the $^{235}$U antineutrino spectrum at a
compact research reactor. A search for antineutrino oscillations over
meter-long baselines probes the hypothesis of sterile neutrinos in the
appropriate mass region and has discovery potential to new physics. A
measurement of the flux and spectrum can inform our understanding of
antineutrino emission from reactors.  Over distances of some fifty
kilometers the observed oscillated energy spectrum provides a unique
method for the determination of the mass hierarchy along with
precision measurements of neutrino mixing parameters.

\subsubsection{Reactors and Experimental Facilities}
The US has several reactor facilities that are well-suited for
short-baseline experiments at distances of $\mathcal{O}$(10~m). The
NIST, HFIR, and ATR reactors in the US operate at 20--250~MW with
highly enriched $^{235}$U, provide easy access and technical support
for a world-leading short-baseline reactor neutrino experiment. Other
compact, high-power sources for antineutrinos such as a naval
reactor are being explored. Through international collaboration the US
neutrino physics community also has an exciting opportunity to
participate in a medium baseline experiment overseas near the Taishang
and Yangjiang reactors in China.


\subsubsection{Small-Scale \& Short-Baseline: Sterile Neutrinos and Reactor Spectrum}
{\em A domestic short-baseline experiment designed to resolve the
reactor neutrino anomaly through oscillation and spectral measurements
has the potential to discover new physics in the next 3--5 years at
modest costs of \$3--4M.}  Proposed short-baseline reactor
disappearance experiments such as NuLAT~\cite{Lane:2015alq} and
PROSPECT~\cite{Ashenfelter:2013oaa} are complementary to the FNAL
short-baseline program focusing on appearance measurements, and will
answer both the question of short-baseline oscillation and our
understanding of the reactor antineutrino spectrum. The proposed
projects have deployed test detectors at NIST and HFIR, are ready to
proceed with full design and construction, and provide an opportunity
for world-leading, high-impact science at modest costs. In a
technically limited schedule, data taking can begin in 2016 with first
physics results in 2017. Given US experience and available reactor
facilities, the US is in an excellent position to host and lead a
short-baseline reactor experiment. The experiments offer opportunities
for international collaboration with Canada, China, and Europe.
Several other efforts reactor experiments worldwide are at various
stages of development. This includes SOLID, STEREO, DANSS, and
Neutrino-4. Experiments proposed in the US aim to optimize physics
sensitivity to both the reactor spectrum and neutrino oscillations,
proceed in a phased approach, and provide comprehensive systematic
control through novel scintillator and detector technologies, movable
detectors, and extensive background control. Detectors proposed in the
US offer the highest energy resolution of $\sim$4.5\% which provides
unmatched capability in the measurement of the reactor spectrum. This
comprehensive and phased approach will maximize the discovery
potential, limit technical risk, and provide flexibility to respond to
future discoveries.  Timely execution is critical to guarantee the
highest impact and facilitate European participation in a US
experiment.

\subsubsection{Mid-Scale \& Medium-Baseline: Mass Hierarchy and Oscillation Parameters}
 {\em Medium-baseline experiments aim to determine the neutrino mass
hierarchy without matter effect and precisely measure $\theta_{12}$,
$\Delta m^2_{21}$, and $\Delta m^2_{32}$. over the next 7--10
years~\cite{Kettell:2013eos}. A US contribution can make a critical
impact to one of these overseas experiments.} The JUNO experiment in
China and RENO-50 in Korea are designed to determine the neutrino mass
hierarchy and precisely measure oscillation parameters. In particular,
the JUNO experiment, hosted by China and with substantial European
contributions, offers strong leveraging of the successful US-China
collaboration. The US is well-positioned to make an important
contribution to JUNO.  Near-term R\&D of a few \$M can help define US
scope focused on the calibration system with an eventual US JUNO
contribution of order \$20M over 5--8 years.  A reactor experiment at
medium baselines represents an excellent opportunity to continue the
long-standing US-China collaboration and ensures US leadership in the
determination of the mass hierarchy.

\subsubsection{Synergies} 
{\em Measurements of reactor neutrinos are also relevant to nuclear
physics and applied reactor safeguards. The proposed experiments will
continue the long-standing US expertise in this area.} Short-baseline
experiments provide opportunities for detector R\&D into technologies
such as novel neutron-sensitive scintillators and highly segmented
detectors. This work leverages R\&D and LDRD support from several
sources including DOE HEP, NP, and NSF as well as NIST.  Large reactor
antineutrino detectors offer synergistic opportunities for
geo-neutrino physics and astrophysics. Many of the theoretical and
experimental challenges are common across these fields and reactor
neutrino measurements have the potential to uniquely inform these
communities.

\newpage
\subsection{Source, Cyclotron and Meson Decay at Rest Neutrinos}
\label{sec:SourceCyclotronDAR}

Isotope and meson decay-at-rest (DAR) processes can be used to provide
very high-intensity sources of neutrinos and antineutrinos with
energies spanning a few MeV to a few hundred MeV. This energy range
is ideal for a number of physics measurements including sterile
neutrino searches with baselines of 10--100~meters, searches for and
the possible discovery of coherent elastic neutrino-nucleus
scattering (CEvNS), and neutrino cross section measurements relevant
for astrophysical processes such as supernova explosions. There are
trade-offs in cost, schedule, and sensitivity so a diverse program of
possible technologies and setups can cover the wide range of possible
physics opportunities. High-activity radioactive sources can be
produced at reasonable cost and, when coupled with large,
low-background detectors, can have good sensitivity to electron
neutrino and antineutrino oscillations to sterile neutrinos. Very high
rates of radioactive isotopes that produce higher energy neutrinos can
be continuously produced by high-intensity cyclotrons and lead to
conclusive studies of sterile neutrino oscillations. Spallation
neutron sources such as the SNS and the JPARC-MLF facilities are
copious sources of neutrinos from meson DAR in the spallation
production dumps. Small, fairly low-cost detectors with good low
energy capabilities can be used at these facilities to search for
the elusive CEvNS process. These facilities could also host very
sensitive sterile neutrino oscillation experiments using large
detectors in the 50--1000~ton scale. The sections below outline the
features of experiments using these sources including
information on cost, timescale, and physics coverage.

\subsubsection{Small-scale Experiments}

\noindent{\bf Radioactive Source Experiments}

\noindent Radioactive source experiments could be a cost effective way
to investigate electron antineutrino disappearance in the region of
the reactor anomaly. The SOX program~\cite{bib:SOX} will use
high-intensity radioactive sources of neutrinos or antineutrinos to
investigate disappearance oscillations. The initial data run would
use sources placed below the Borexino detector. Details of the cerium
source production are described in~\cite{bib:CL}. This phase is
expected to have physics results within 5 years (cerium run will be
finished by the end of 2017, followed by a chromium run that should be
finished by mid 2018). The physics results are expected to probe at
95\% C.L. the entire Reactor Antineutrino Anomaly region. This phase
will also provide important R\&D for future upgrades relevant to SOX
and the US $^{51}$Cr program on LZ, SNO+, RICOCHET and possibly
others. This initial phase would have a cost in
the \$2--3M range and fit into the proposed FOA category. 
Future upgrades could include higher intensity sources and deployment
of the sources within the detector to get enhanced oscillation
parameter space coverage.

\noindent{\bf JPARC-MLF Pion/Kaon Decay-at-Rest Experiment}

\noindent JPARC-E56~\cite{JPARC_P_56} will directly probe the LSND
anomaly with $\bar\nu_e$ appearance using a 50~ton Gd-doped liquid
scintillator detectors at the JPARC-MLF (1~MW) 3~GeV spallation
neutron facility. First data is expected in the next 2--3 years. The
experiment will provide competitive (95\% C.L. coverage), but not
definitive, sensitivity to the LSND allowed region. The project is of
modest scale with Japanese costs at the \$5M level, but has the
potential for upgrades and additional detector modules. The main US
contribution to JPARC-E56 will be the dilute Gd-loaded liquid
scintillator, which represents an important technological R\&D step in
producing doped scintillator for detecting both Cherenkov and
scintillation signals. This effort will fit well into the FOA scope
with an estimated cost of \$1.5M.

The JPARC-MLF beam also uniquely allows the possibility to study kaon
decay-at-rest (KDAR)~\cite{KDAR,KDAR1} muon neutrinos 
and related physics for the first time with an expected sample of over 10$^5$ muon
neutrino charged current events. The KDAR muon neutrinos are
mono-energetic with an energy of 236~MeV. As the only relevant
known-energy muon neutrino above the charged current threshold, this
unique neutrino can be used for studying short baseline oscillations
indicative of a sterile flavor, neutrino cross sections relevant for
future CP violation searches, and nuclear physics with a known-energy,
weak-interaction-only probe.

\noindent{\bf Coherent Elastic Neutrino-Nucleus Scattering Experiments}

\noindent Coherent Elastic Neutrino-Nucleus Scattering (CEvNS) is an
unambiguous prediction of the Standard Model. Recent advances in
detector design now put this so-far elusive prize within reach. Such a
measurement will open the door to a host of new ways to better
understand neutrino properties, and to search for new physics. CEvNS
represents an eventually dominant background for dark matter
detection, and its measurement will demonstrate dark matter detector
response. As a neutral current process, it will be a new tool for
sterile neutrino oscillation experiments.

The COHERENT experiment~\cite{Akimov:2013yow} will search for CEvNS at the
SNS with three detector targets (CsI, Xe, Ge). The 1.4~MW SNS has a
flux of 3.3$\times$10$^{7}$ $\nu$ cm$^{-2}$ s$^{-1}$ at 20~m with a
clean pion DAR spectrum. A neutron measurement campaign has identified
several suitable deployment sites within 30 m of the source. Phase 1
of the experiment is likely to produce initial results within the
year, and would have a cost of \$2--3M, fitting well into the proposed
FOA category.

The CENNS experiment~\cite{Brice:2013fwa} will search for CEvNS at the
BNB with a redeployment of the MINICLEAN detector to Fermilab. The
32~kW BNB has a flux of 5$\times$10$^{5}$ $\nu$ cm$^{-2}$ s$^{-1}$ at
20~m. Measurements indicate that backgrounds from neutrons are
manageable with sufficient shielding, in a green field site. The first
results are expected after 2018, and are coupled to the physics
program of MINICLEAN. The expected cost is \$2M, which fits well into
the proposed FOA category.

\subsubsection{Mid-scale Experiments}

\noindent{\bf The IsoDAR Isotope Decay-at-Rest Experiment}

\noindent The IsoDAR experiment~\cite{IsoDAR,IsoDARJuno} will make a
highly definitive investigation of electron antineutrino disappearance
in the reactor anomaly region and make precision electroweak
measurements searching for neutrino non-standard interactions. The
experiment uses a very high intensity $^8$Li antineutrino source placed
near a large scintillator detector such as KamLAND or JUNO. IsoDAR
can also be coupled with WATCHMAN and provide an important component
of the WATCHMAN physics program. The cost of IsoDAR is estimated to
be \$30M so it is a mid-scale project. At this point, the development
of the IsoDAR cyclotron and neutrino source needs engineering R\&D
support at about the \$1M level to complete prototypes and prepare a
Conceptual Design Report to be submitted to the agencies.

\noindent{\bf The OscSNS Pion Decay-at-Rest Experiment}

\noindent OscSNS~\cite{OscSNS} has the capability to make a definitive
search for electron antineutrino appearance using a pion DAR beam with
much lower uncertainties than LSND. The cost is at the \$12M scale for
civil construction and \$8M for a new 800~ton liquid scintillator
detector with a start date 3 years after initiation. With the high
neutrino rate at SNS, the experiment can cover the full LSND signal
region in 2 years with some capability to see oscillatory behavior for
$\Delta$m$^2 > 1$~eV$^2$. The successful support by DOE NP for
infrastructure development at the SNS related to the fundamental
physics neutron beamline and support facility can serve as a model for
future infrastructure development related to OscSNS and other neutrino
efforts supported by DOE HEP.

\subsubsection{Cross Section Measurements using DAR Neutrino Sources}

\noindent DAR neutrino sources can be used to measure a number of
important neutrino interaction cross sections. Neutrino-nucleus
cross-section measurements on various targets are inputs to supernova
modeling and for understanding supernova detectors. Specifically, the
CENNS, CAPTAIN-BNB~\cite{Berns:2013usa}, JPARC-MLF, and COHERENT (Ge,
CsI, Xe) at SNS experiments will address many of these cross-section
measurements. For example, ongoing measurements of the
neutrino-induced neutron background for COHERENT play an important
role in r-process nucleosynthesis, as well as in the HALO SNe
experiment~\cite{halo}. To make these measurements one needs to
understand the flux spectrum, the detector characteristics and the
backgrounds. Neutron backgrounds are the most important and need to be
addressed by location or shielding.

\newpage
\subsection{Neutrino Detector R\&D}
\label{sec:RandD}

\subsubsection{Water and Liquid Scintillator}

Development of new scintillator materials and doping agents has proven
critical to the advancement of neutrino detector design. Further
development of these materials is a critical step for future
experiments; supporting this effort should be a high priority in the
intermediate program. This program includes target development and
characterization, including: light yield and timing measurements at
low and high energy, energy nonlinearity, and attenuation
measurements.

The newly-developed water based liquid scintillator (WbLS) could
enable a massive detector with a broad physics program at relatively
low cost. A particularly nice feature is the potential to separate
fast, directional Cherenkov light from the slower yet far more
abundant isotropic scintillation light. Should this be achieved, this
would enable astonishing advances in signal identification and
background rejection capabilities via particle identification,
resulting in vast improvements in physics reach. This potential
capability should be explored via both optimization of the WbLS target
--- by modifying the LS fraction and thus relative magnitudes of the
Cherenkov and scintillation components, by use of various additives to
delay the scintillation light, or wavelength shifters to minimize
absorption/reemission of Cherenkov light --- and alternate photon
detection methods. The ability to reconstruct event energy and
direction in (Wb)LS needs to be demonstrated both theoretically (in
simulation) and in practice (in small scale experiments).

Large water Cherenkov and (water-based) scintillator detectors require
very high purity target liquids.  Purification techniques for water
are well understood in industry and need no R\&D, but further
development is required for (Wb)LS purification. At the same time, a
program to determine compatibility of construction materials with
(Wb)LS must exist for future detectors.

Isotope loading in traditional liquid scintillator, Gadolinium doping
in water detectors, and the potential to load metallic isotopes in
WbLS broaden the potential physics program and significantly enhance
the sensitivity of future experiments. Techniques to load isotope
while maintaining the optical purity of the target should continue to
be developed.

A driving cost and critical performance factor in large-scale water or
scintillator detectors is the photomultiplier tubes. R\&D to produce
low cost, large area, ultra-fast photon detectors is important for the
neutrino community. Correspondingly fast, high precision readout will
be critical to take advantage of developments in photon detector
technology.

Water-based detectors (including WbLS) have the advantage of a low
cost detector medium allowing very large-scale experiments. Future
experiments will be limited by the cost and excavation techniques for
the cavern needed to house the experiments. R\&D to find lower-cost
construction methods, including PMT deployment and readout techniques,
can facilitate next-generation neutrino detectors.

Several projects are underway that can address these topics. These
range from bench-top scale development and characterization of
newly-developed WbLS and LS for next-generation large-scale detectors,
primarily at BNL but also at U. Chicago, U. Penn, LBNL, Iowa State,
and MIT, to full-scale projects such as EGADS (Gd loading), ANNIE
(fast timing), WATCHMAN phase II (WbLS deployment, fast timing), SNO+
(Te loading), and CHIPS (large-scale construction). This collaborative
effort is supported by DOE-HEP, DOE-NP, NSF and LDRD. Ultimately such
projects will inform the design of a massive future detector such as
the proposed \textsc{Theia} experiment.

\subsubsection{Liquid Argon}

Several ongoing and proposed experimental efforts will provide R\&D
that will substantially improve understanding of LArTPC performance or
potentially expand the capabilities of the ELBNF experiment. Test beam
measurements are especially important as they provide critical data
for improving the detector model and understanding of systematic
uncertainties.  The majority of these experiments will use the test
beam facilities at FNAL or CERN or the neutrino beams at FNAL. Support
for these efforts can be a combination of R\&D funding at FNAL, SBN or
ELBNF project funding, funding from an intermediate neutrino program
or other agency or laboratory funding.  Careful evaluation and
prioritization of these experiments by the FNAL PAC including an
evaluation of impact of R\&D by these experiments is expected, and
should provide important guidance to the selection process. However,
the substantially larger funding available to the FNAL projects should
also be taken into account. Only one experiment that will provide
critical R\&D needed for the ELBNF program is outside the FNAL
program: the neutron cross section measurements proposed by the
CAPTAIN experiment are necessary to understand the detector response
and energy resolution.

The following factors should be taken into account when evaluating the
impact different experiments could have on the long range program:
\begin{itemize}

\item A comprehensive test beam program must be performed to
  characterize present and future LArTPCs. This is necessary to
  calibrate the detector response of existing and future LAr detectors
  and to verify systematic error estimates. This program should
  include electromagnetic and hadronic showers measurements, neutron
  cross section measurements, and energy deposition measurements with
  different charged particle beams at appropriate
  energies. Experiments which could contribute to this are LArIAT,
  CAPTAIN, and the CERN neutrino platform experiments.

\item R\&D on the generation and breakdown of high voltage will reduce
  the risk to future LAr detectors and could lead to more monolithic
  and lower cost detector designs based on longer drifts. The causes
  of HV breakdown in LAr are not well understood. If the process for
  HV discharge in LAr is better understood then detectors could be
  designed for higher voltages (if the electron lifetime is
  sufficiently large). This could lead to larger, cheaper detectors
  and could enable dual-phase style detectors. R\&D on LAr
  feedthroughs above 100~kV is needed as manufacture of LAr feedthrus
  for voltages above 100 kV has only been achieved successfully by a
  small number of groups.

\item The processes for contamination generation and transport inside
  large liquid argon detectors are not well understood. Better
  modeling of the sources and migration of contaminants in large
  cryogenics systems will aid future detector design.

\item R\&D that improves the understanding of the generation and
  propagation of both light and charge in large LArTPCs will improve
  the detector model for ELBNF.
 
\item Present photon detector designs capture a very small fraction of
  the scintillation light generated in large LArTPCs. Detectors with
  better light collection efficiency should be developed.

\item Development of cold electronics for LArTPCs is
  critical. Advanced designs for cold preamps and ADCs exist but a
  control chip is only in early stages of development. Development of
  electronics to read out large arrays of SiPMs is necessary.

\end{itemize}

\newpage

\subsection{Neutrino Theory}
\label{sec:Theory}

During the workshop on Intermediate Neutrino Program held at Brookhaven National Laboratory in February 2015, the neutrino theory community
organized a vibrant session which highlighted the physics goals that can be achieved in the near term future and the role that  theory plays
in achieving these goals.  This section summarizes the consensus that emerged from these discussions. We include specific recommendations which should help the funding agencies in considering support for the neutrino program through investments in theory; including, but not limited to any potential FOA in this area.

Theory plays an essential role in making advances in neutrino physics.
Theory input is necessary in almost all facets of experimental neutrino physics.  Theoretical tools are required
to interpret cross section measurements, to formulate oscillation paradigm with 3 or more neutrinos, to seek
the underlying neutrino mass generation mechanism, to connect experiments with cosmology and astrophysics,
and to infer the fundamental properties of neutrinos.

A relatively small investment in neutrino theory in the intermediate time scale will provide huge dividends that would enrich the
community as a whole.  It would result in timely development of theoretical models for
neutrino-nucleus cross sections, phenomenological studies on
the impact of cross section uncertainties,
theoretical cross checks of the 3 neutrino oscillation paradigm with tools such as global fits,
model-building advances to understand large mixings and possible light sterile
neutrinos, and potentially unravel the underlying symmetry that plays a role in neutrino mass generation.
{\it The neutrino theory community thus recommends support for neutrino theory and that the language of any potential FOA for the intermediate neutrino program be broad
enough to allow potential theory proposals.}


The theory perspective on the physics goals that can be achieved in the near term future are highlighted below.

\noindent {\bf (A) Fundamental neutrino properties:}

{\bf (1)} {\it Dirac versus Majorana nature of the neutrino:}  Improved searches for neutrinoless double
beta decay are the best bet to address this fundamental question.  If neutrinos obey an inverted mass spectrum
observable signals may be within reach in such experiments in the near future.  Searches for neutrinoless double 
beta decay should continue without waiting for the mass hierarchy measurement, as there may be surprises here, 
such as lepton number violation mediated by TeV scale particles.
{\bf (2)} {\it Direct neutrino mass measurement}: Determining the absolute neutrino mass scale in
tritium beta decay experiments would provide deep insight into the origin of neutrino masses.
{\bf (3)} {\it Sensitivity to mass hierarchy and possibly CP violation:}  Any progress towards measuring the neutrino mass
hierarchy before ELBNF would be highly desirable. In particular, knowing the hierarchy is crucial for the possibility of
obtaining a hint for the value of the leptonic CP violating phase prior to ELBNF.
{\bf (4)}  {\it Consistency checks of the three neutrino oscillation paradigm:}  This requires a variety
experimental information including: (a) Improved knowledge of neutrino oscillation parameters from solar, atmospheric, accelerator and reactor
neutrino experiments; and (b)
Essential information on neutrino interaction rates from experiments, aided by theory.

\noindent {\bf (B) Neutrino interactions:}

Knowing the interaction rates is crucial for addressing many of the important questions in neutrino physics.  For example, 
Over much of the available parameter space, discovery of leptonic CP violation at ELBNF will require, as-yet unachieved, percent-level control over $\nu_e$ appearance signals. A dominant source of uncertainty on this signal is due to the modeling of neutrino interactions with the target nucleus in the near and far detectors. Relating the fundamental quark-level interactions of the neutrino to the complete nuclear response is a difficult field theory problem, involving both particle and nuclear physics.  Unlike in the collider physics community, where there is vibrant interactions between researchers in the domains of (i) detector modeling and event simulation at the hadron level, (ii) perturbative QCD analysis at the parton level and (iii) model building and theoretical interpretation, presently the situation is very different in the analysis of signals at accelerator based neutrino experiments. The analogs of (i) and (ii) above are both relegated to the nuclear physics community. This has two unfortunate outcomes. First, many tools in the particle theorist's toolkit are not brought to bear on these problems. Second, there is an institutional barrier to communication between those researchers studying neutrino models, and those involved in understanding the experimental analysis of signals and backgrounds.
Collaborative efforts involving HEP theorists, nuclear theorists and neutrino experimenters can uplift this area of research to a level comparable to the one seen in collider physics today.


\noindent {\bf (C) Short baseline anomalies and sterile neutrinos:}

{\bf (1)} {\it Understanding anomalies seen in short baseline experiments}:  Unambiguous resolution
in terms of oscillations would require seeing $L/E$ dependence in new/upgraded experiments.
{\bf (2)} {\it Existence of sterile neutrino}:  Discovery of new sterile states in neutrino oscillation
experiments attempting to resolve short baseline anomalies will be foundational.
{\bf (3)} {\it Nonstandard neutrino interactions}:  If discovered, these effects would invalidate the
three neutrino oscillation paradigm, and hint at new physics beyond neutrino masses.

\noindent {\bf (D) Neutrinos in astrophysics and cosmology:}

{\bf (1)}  {\it Supernova neutrinos}:  Anticipating neutrinos from supernova explosions in the near term future, we feel that
investment in understanding the complex dynamics of collective neutrino oscillation is necessary, and that this should happen now,
as it could influence detector design decisions in the next several years.
{\bf (2)} {\it High energy astrophysical neutrinos}: IceCube and its upgrade will tell us more about
the origin of very high energy (PeV scale) astrophysical neutrinos.  Energy spectrum, directional information,
and flavor composition of these events can help us understand the astrophysical sources as well as neutrino properties.
{\bf (3)} {\it Neutrinos in cosmology}: Although indirect, neutrino masses inferred from cosmology
would provide complementary information, and at the same time also test standard cosmological models.

\noindent {\bf (E) Underlying symmetries behind neutrino masses:}

{\bf (1)} {\it Neutrino masses and physics beyond the Standard Model}: What can neutrino experiments teach
us about the underlying symmetries of the theory that generates neutrino masses?  Experiments in the intermediate
time scale can lead to progress in this very basic question.
{\bf (2)} {\it Nucleon decay}:  Ongoing large underground detectors (Super-Kamiokande) which are sensitive to neutrino
oscillation physics continue to be also sensitive to nucleon decay. Discovery of nucleon decay would be monumental.
{\bf (3)} {\it Exotic neutrino properties}:  Information on neutrino properties such as its magnetic moment,
decay lifetime, possible violations of Lorentz invariance and/or CPT invariance, and its interactions with
the dark matter sector would be valuable.  Even if
not found, neutrinos can provide some of the best tests of these fundamental symmetries.


%
\newpage
\bibliographystyle{winp}
\bibliography{WINP}

\newpage
\section{WINP Agenda}
\label{sec:Agenda}
\url{http://www.bnl.gov/winp/}
\begin{figure}[h] 
\includegraphics[height=0.9\textheight]{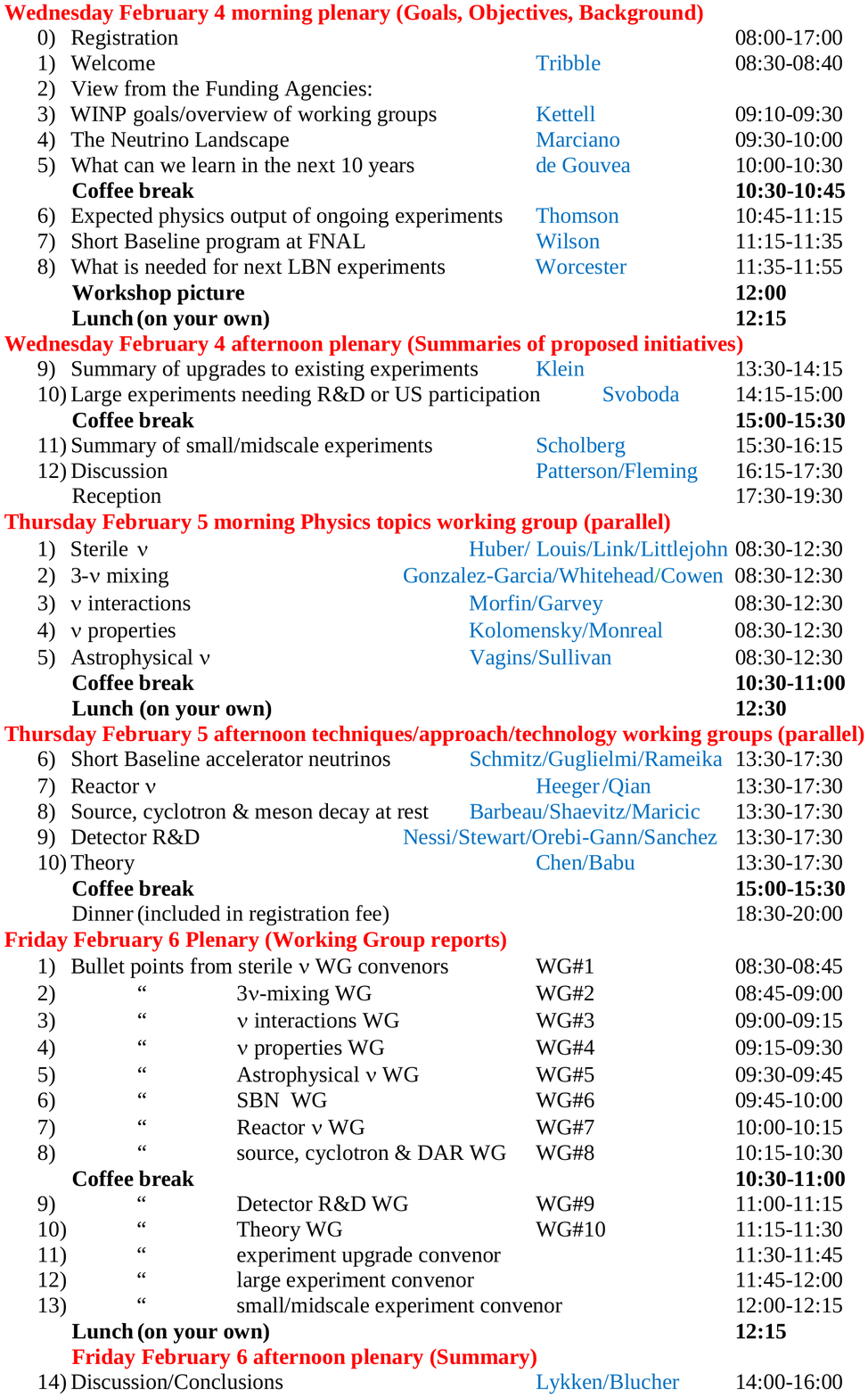}
\end{figure}

\newpage
\section{Experimental Questionaire Responses}
\label{sec:Experiments}

Responses from experiments to the WINP questionaire follow.\\
(see also \url{https://indico.bnl.gov/conferenceDisplay.py?confId=918})

\begin{multicols}{2}
\begin{enumerate}
  \item ANNIE
  \item ARA
  \item ASDC (\textsc{Theia})
  \item CAPTAIN
  \item CENNS
  \item CeSOX 
  \item CHIPS
  \item COHERENT
  \item Cr51
  \item CUORE
  \item DAEdALUS
  \item DayaBay
  \item ELBNF
  \item Hyper-K
  \item IceCube
  \item IsoDAR
  \item Jinping
  \item JPARC56
  \item JUNO
  \item KamLAND
  \item KATRIN
  \item LAr35ton
  \item LAr-CERN-prototype
  \item LArIAT
  \item MINERvA
  \item MINOS
  \item NESSiE
  \item nEXO
  \item NEXT
  \item NuLAT
  \item NuPRISM
  \item OscSNS
  \item PINGU
  \item Project-8
  \item PROSPECT
  \item RICOCHET
  \item SBN-ICARUS
  \item SBN-LAr1-ND
  \item SNO+
  \item Super-Kamiokande
  \item Super-NEMO-Demonstrator
  \item US-NA61
  \item WATCHMAN
\end{enumerate}
\end{multicols}

\end{document}